\documentclass[reprint,amsmath,amssymb,nofootinbib,aps,prc]{revtex4-1}
\usepackage{caption}
\usepackage{subfigure,dcolumn}
\usepackage{amsmath}
\usepackage{epstopdf}
\usepackage{mhchem}
\usepackage{upgreek}
\usepackage{bm}
\usepackage{color}
\usepackage{graphicx}
\usepackage{lmodern}
\usepackage{mathtools}
\usepackage{multirow}
\usepackage{makecell}
\usepackage{textcomp}
\usepackage{ulem}
\usepackage{xurl}
\usepackage{xspace}

\newcommand{\pt} {\ensuremath{p_{\rm T}}\xspace}
\newcommand{\vtwo} {\ensuremath{v_2}\xspace}
\newcommand{\vtwosp} {\ensuremath{\vtwo \{\rm SP\}}\xspace}
\newcommand{\vtwoep} {\ensuremath{\vtwo \{\rm EP\}}\xspace}
\newcommand{\vtwotwo} {\ensuremath{\vtwo \{2\}}\xspace}
\newcommand{\vtwofour} {\ensuremath{\vtwo \{4\}}\xspace}
\newcommand{\etwo} {\ensuremath{\varepsilon_2}\xspace}
\newcommand{\etworp} {\ensuremath{\varepsilon_2 \{\rm RP\}}\xspace}
\newcommand{\etwotwo} {\ensuremath{\varepsilon_2 \{2\}}\xspace}
\newcommand{\etwofour} {\ensuremath{\varepsilon_2 \{4\}}\xspace}

\begin{document}

\title{Particle species dependence of elliptic flow fluctuations in Pb–Pb collisions at LHC energies in a multiphase transport model}

\author{Jie Wan$^{a,b}$}
\affiliation{$^a$Key Laboratory of Nuclear Physics and Ion-Beam Application (MOE), Institute of Modern Physics, Fudan University, Shanghai 200433, China}
\affiliation{$^b$Shanghai Research Center for Theoretical Nuclear Physics, NSFC and Fudan University, Shanghai 200438, China}

\author{Chun-Zheng Wang$^{a,b}$}
\affiliation{$^a$Key Laboratory of Nuclear Physics and Ion-Beam Application (MOE), Institute of Modern Physics, Fudan University, Shanghai 200433, China}
\affiliation{$^b$Shanghai Research Center for Theoretical Nuclear Physics, NSFC and Fudan University, Shanghai 200438, China}

\author{Yu-Gang Ma$^{a,b}$}
\email{mayugang@fudan.edu.cn}
\affiliation{$^a$Key Laboratory of Nuclear Physics and Ion-Beam Application (MOE), Institute of Modern Physics, Fudan University, Shanghai 200433, China}
\affiliation{$^b$Shanghai Research Center for Theoretical Nuclear Physics, NSFC and Fudan University, Shanghai 200438, China}

\author{Qi-Ye Shou$^{a,b}$}
\email{shouqiye@fudan.edu.cn}
\affiliation{$^a$Key Laboratory of Nuclear Physics and Ion-Beam Application (MOE), Institute of Modern Physics, Fudan University, Shanghai 200433, China}
\affiliation{$^b$Shanghai Research Center for Theoretical Nuclear Physics, NSFC and Fudan University, Shanghai 200438, China}

\author{Song Zhang$^{a,b}$}
\affiliation{$^a$Key Laboratory of Nuclear Physics and Ion-Beam Application (MOE), Institute of Modern Physics, Fudan University, Shanghai 200433, China}
\affiliation{$^b$Shanghai Research Center for Theoretical Nuclear Physics, NSFC and Fudan University, Shanghai 200438, China}

\begin{abstract}
The fluctuations of elliptic flow (\vtwo) in relativistic heavy-ion collisions offer a powerful tool to probe the collective behavior and transport properties of the quark-gluon plasma (QGP). The dependence of these fluctuations on particle species further sheds light on the hadronization mechanism. At LHC energies, the ALICE experiment has measured $v_2$ fluctuations for charged pions, kaons, and (anti-)protons via the ratio of \vtwo measured with respect to the spectator plane (\vtwosp) and from the four-particle cumulants (\vtwofour). However, the observed dependencies on transverse momentum and particle type remain not fully understood. In this study, we perform a phenomenological investigation using a multiphase transport (AMPT) model, which allows us to trace the full evolution of flow fluctuations intertwined with the quark coalescence. The results qualitatively reproduce the ALICE measurements and offer deeper insights into the transport dynamics and hadronization of the QGP.

\end{abstract}

\maketitle

\section{Introduction}

The formation and the evolution of the color-deconfined strongly interacting matter, quark-gluon plasma (QGP) have been studied via relativistic heavy-ion collisions for decades~\cite{alice_qgp, nst_alice, nst_star}. As one of the most important probes, elliptic flow (\vtwo), defined as the second-order Fourier component of the azimuthal distribution of emitted particles ($\phi$) with respect to the collision symmetry plane ($\Psi$), $\vtwo \equiv \langle \rm cos2(\phi - \Psi) \rangle$, plays an essential role in revealing the transport and hydrodynamic properties of the QGP~\cite{flow_review1}. It is believed that such an anisotropic collective motion stems from the elliptical spatial geometry of the shape of the overlap region of two colliding nuclei, the latter commonly characterized by the eccentricity $\varepsilon_2$. Ideally, hydrodynamics suggest that \vtwo should be linearly proportional to the $\varepsilon_2$: $\vtwo \propto \etwo$. Realistically, however, due to the motion of nucleons and the quantum mechanical nature, initial positions of nucleons within the participant zone fluctuates from collision to collision, causing fluctuations of $\varepsilon_2$ and deviations between the genuine reaction plane (RP) and the participant plane (PP)~\cite{ecc1, ecc2, ecc3}. Therefore, the anisotropic flow of final stage particles are correspondingly fluctuates event by event. The comparison of the measured \vtwo coefficients and their respective eccentricities is crucial for investigating the evolution of the heavy-ion collisions, in particular, the QGP viscosity and the hadronization process. 

Flow fluctuations ($\sigma_{v_2}$) manifest themselves through different methods of calculating \vtwo: \vtwo w.r.t. the reaction plane $v_2\{\rm RP\}$, \vtwo calculated through 2 particle cumulants $v_2\{2\}$, and through 4 particle cumulants $v_2\{4\}$~\cite{flow_ep, flow_cumu1, flow_cumu2}:
\begin{equation}
\begin{aligned}
v_2\{\rm RP\} &= \rm cos2(\phi - \Psi_{RP}), \\
v_2\{2\} &= d_2\{2\} / \sqrt{c_2\{2\}}, \\
v_2\{4\} &= -d_2\{4\} / (-c_2\{4\})^{3/4}, \\
\end{aligned}
\label{Eq:v2}
\end{equation}
where $d$ and $c$ denote the cumulants for the differential and reference flow, respectively.
Disregarding non-flow effects, $v_2\{\rm RP\}$ reflects the genuine \vtwo by definition, while $v_2\{2\}$ and $v_2\{4\}$ are influenced by flow fluctuations. Early studies~\cite{ecc2} suggest:
\begin{equation}
\begin{aligned}
v_2\{2\} &\simeq  \langle v_2 \rangle + \sigma_{v_2}^2 / 2\langle v_2 \rangle, \\
v_2\{4\} &\simeq \langle v_2 \rangle - \sigma_{v_2}^2 / 2\langle v_2 \rangle. \\
\end{aligned}
\label{Eq:fluc}
\end{equation}
As \vtwo in high energy heavy-ion collisions is typically positive, the hierarchy $v_2\{2\} > v_2 > v_2\{4\}$ holds, and the deviation among them can be used to quantify flow fluctuations.
The initial eccentricity $\varepsilon_2$ can be similarly characterized through~\cite{ecc1, ecc2, ecc3}:
\begin{equation}
\begin{aligned}
\varepsilon_2 \{\rm RP\} &= \langle \rm y-x \rangle / \langle \rm y+x \rangle \\
\varepsilon_2\{2\} &= \sqrt{\langle \varepsilon_{\rm part}^2 \rangle} \\
\varepsilon_2\{4\} &= (2\langle\varepsilon_{\rm part}^2 \rangle^2 - \langle\varepsilon_{\rm part}^4 \rangle)^{1/4} \\
\end{aligned}
\label{Eq:e2}
\end{equation}
where $(x,y)$ represents the position of a particle of interest in the coordinate system, and $\varepsilon_{\rm part}$ denotes the participant eccentricity.
As per Refs~\cite{ecc1, fluc1, star_flow_fluc1}, the ratios $\vtwo\{\rm RP\} / \etwo\{\rm RP\} \simeq \vtwotwo / \etwotwo \simeq \vtwofour / \etwofour$, provide a critical constraint for determining QGP transport properties.

To test these relations experimentally, the ALICE collaboration has measured \vtwo relative to the neutron spectator plane \vtwosp with inclusive hadrons~\cite{alice_sp_flow, alice_xe_flow}. Since the genuine RP is experimentally inaccessible, the spectator planes (SP) determined by spectator nucleons and fragments serves as a practical proxy.
The measured \vtwosp for charged hadrons reveals a centrality dependent deviation between the ratio \vtwosp/\vtwofour and the eccentricity ratio $\etwo\{\rm RP\} / \etwofour$ predicted by the $\rm T_{R}ENTo$ and EPM models~\cite{trento, epm}. This discrepancy, observed in both central and peripheral collisions, indicates an incomplete theoretical description of initial state fluctuations.

The transverse momentum (\pt) dependent flow is another unique probe for investigating multiple properties of the QGP, including pressure gradients, the degree of thermalization, the equation of state and hadronization mechanism. Specifically, The QGP's radial expansion produces characteristic mass-ordering of \vtwo at low \pt, while quark coalescence during hadronization leads to baryon-meson grouping at intermediate \pt.
In the ALICE measurement~\cite{alice_sp_flow}, a \pt dependence of the ratio \vtwosp/\vtwofour is observed across most centrality intervals, serving as an important experimental input for understanding whether flow fluctuations at low and intermediate \pt originate from common sources.

In addition to inclusive hadrons, the ALICE experiment has recently extended these measurements to identified particles, namely, charged pions ($\pi^\pm$), kaons ($K^\pm$), and (anti-)protons ($p, \bar{p}$)~\cite{alice_sp_pidflow}. A significant particle-species dependence is observed in the ratio \vtwosp/\vtwofour for $\pi$, $K$, and $p$.
Such a dependence is not expected from the linear hydrodynamic response of the QGP to the initial spatial anisotropy. Notably, Ref.~\cite{ampt_flow_fluc1} previously investigated \vtwo fluctuations using \vtwosp\ and \vtwoep but found no particle-species dependence either, leaving this discrepancy an open question.
Therefore, to better understand these experimental findings and the interplay between QGP transport properties and hadronization, we conduct a comprehensive study of \vtwo fluctuations for $\pi$, $K$, and $p$ at LHC energies using a multiphase transport (AMPT) model. which enables us to trace the complete evolution of flow fluctuations and their connection with quark coalescence.
In the following sections, the model and analysis method are presented in Sec.\ref{method}, followed by the main results and discussions in Sec.\ref{results}, and a brief summary in Sec.~\ref{summary}.

\section{Model and analysis method} \label{method}

The hybrid transport model AMPT~\cite{ampt1, ampt2} is widely used for simulating relativistic heavy-ion collisions. Its string melting version, in particular, is known for effectively describing the collective motion of the final state hadrons~\cite{ampt_flow1, ampt_flow2, ampt_flow3, ampt_flow4}. The model consists of four key subroutines which simulate different stages of the collisions in sequence: HIJING for the initial parton condition~\cite{hijing}, ZPC for the partonic evolution~\cite{zpc}, a simple quark coalescence for the hadronization process~\cite{coal} and ART for the hadronic rescatterings and interactions~\cite{art}. Following established parameterizations~\cite{ampt_par1, ampt_par2}, the Lund string fragmentation settings and partonic cross sections in this work are tuned to reproduce the hadron spectra and anisotropic flow at LHC energies. 

To calculate flow observables, we adopt the same definition and method as used in experimental analyses. For \vtwosp, we reconstruct the spectator plane using
\begin{equation}
\begin{aligned}
\Psi_{\rm SP} = \frac{{\rm atan2}(\langle r^2 {\rm sin}2\phi \rangle, \langle r^2 {\rm cos}2\phi \rangle)}{2}, \\
\end{aligned}
\label{Eq:psi}
\end{equation}
where $r$ and $\phi$ denote the spatial coordinate and azimuthal angle of spectator neutrons in the transverse plane, obtained from the initial state of the model. Note that the second order harmonic is used here, though $\Psi_{\rm SP}$ can alternatively be derived from the first harmonic flow vectors~\cite{ampt_flow_fluc1, ampt_cme_sp}. We verify that both methods yield equivalent results, and the calculated $\Psi_{\rm SP}$ is consistent with zero. This aligns with the model’s initial setup, where the impact parameter direction (the genuine $\Psi_{\rm RP}$) is fixed along the $x$-axis. This consistency confirms that, given sufficient resolution, \vtwosp is an ideal observable for probing the genuine \vtwo.
In later sections, we will further examine the differences between the constructed $\Psi_{\rm PP}$ and $\Psi_{\rm SP}$, and their influence on \vtwo fluctuations. The $\Psi_{\rm PP}$ is calculated using the same method as Eq.~(\ref{Eq:psi}), but with initial-state participating partons replacing spectator neutrons, thereby introducing initial fluctuations of eccentricity.

For \vtwotwo and \vtwofour, the standard cumulants method as introduced in Eq.~(\ref{Eq:v2}) is employed. A pseudorapidity ($\eta$) gap of 0.3 between the particle of interest (POI) and reference particle is applied to suppress non-flow effects in the \vtwotwo calculation. All analyzed particles are selected within $|\eta|<$ 1 and transverse momentum range 0.1 $< \pt <$ 4.0 GeV/$c$, matching the experimental acceptance.

\begin{figure}[!htb]
\includegraphics[width=\linewidth]{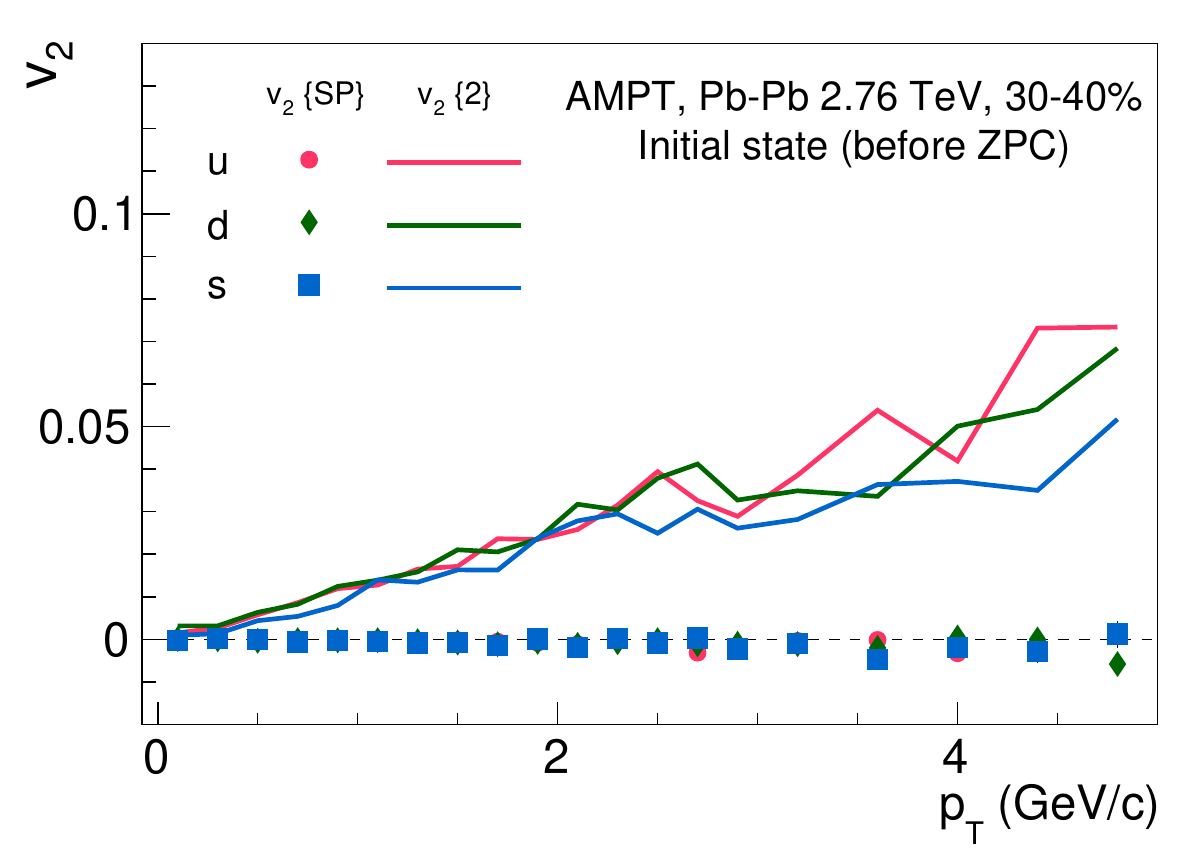}
\caption{(Color online) The \pt dependent $v_2 \{2\}$,  $v_2 \{4\}$ and $v_2 \{\rm SP\}$ for $u$, $d$ and $s$ quarks, in the initial stage before ZPC in 30-40\% AMPT Pb-Pb collisions.}
\label{fig:fig_zpct0}
\end{figure}

\begin{figure*}
\centering
\includegraphics[width=.9\linewidth]{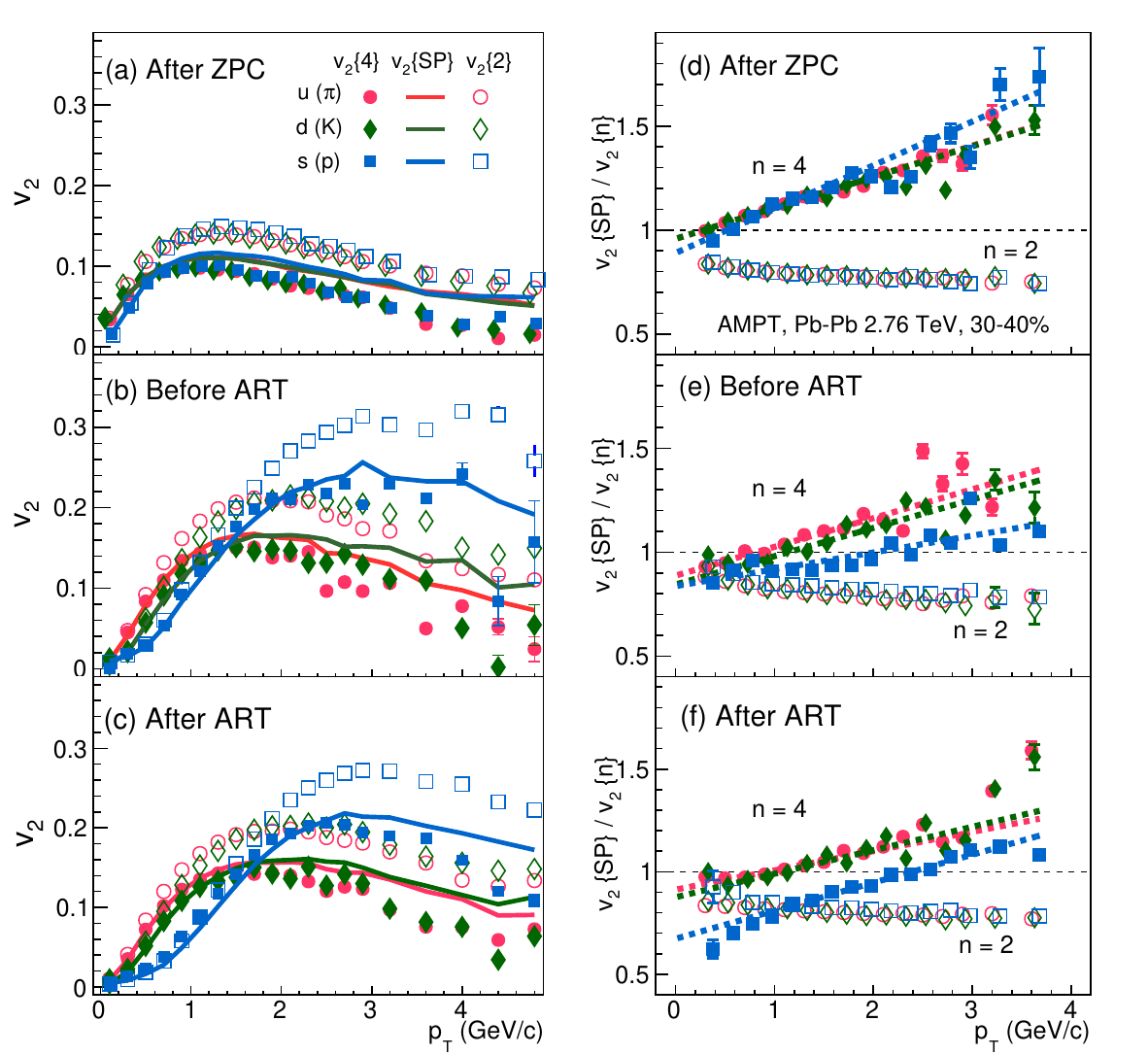}
\caption{(Color online) The \pt dependent $v_2 \{2\}$,  $v_2 \{4\}$ and $v_2 \{\rm SP\}$ and their ratios for $u$, $d$ and $s$ quarks (panels a and d) and for $\pi, K, p$ (panels b, c, e, f), in three evolution stages in 30-40\% AMPT Pb-Pb collisions.}
\label{fig:fig_v2}
\end{figure*}

\section{Results and discussions} \label{results}

\subsection{The \pt dependent $v_2 \{2\}$,  $v_2 \{4\}$ and $v_2 \{\rm SP\}$ for $\pi$, $K$, and $p$}

\begin{figure*}
\centering
\includegraphics[width=.9\linewidth]{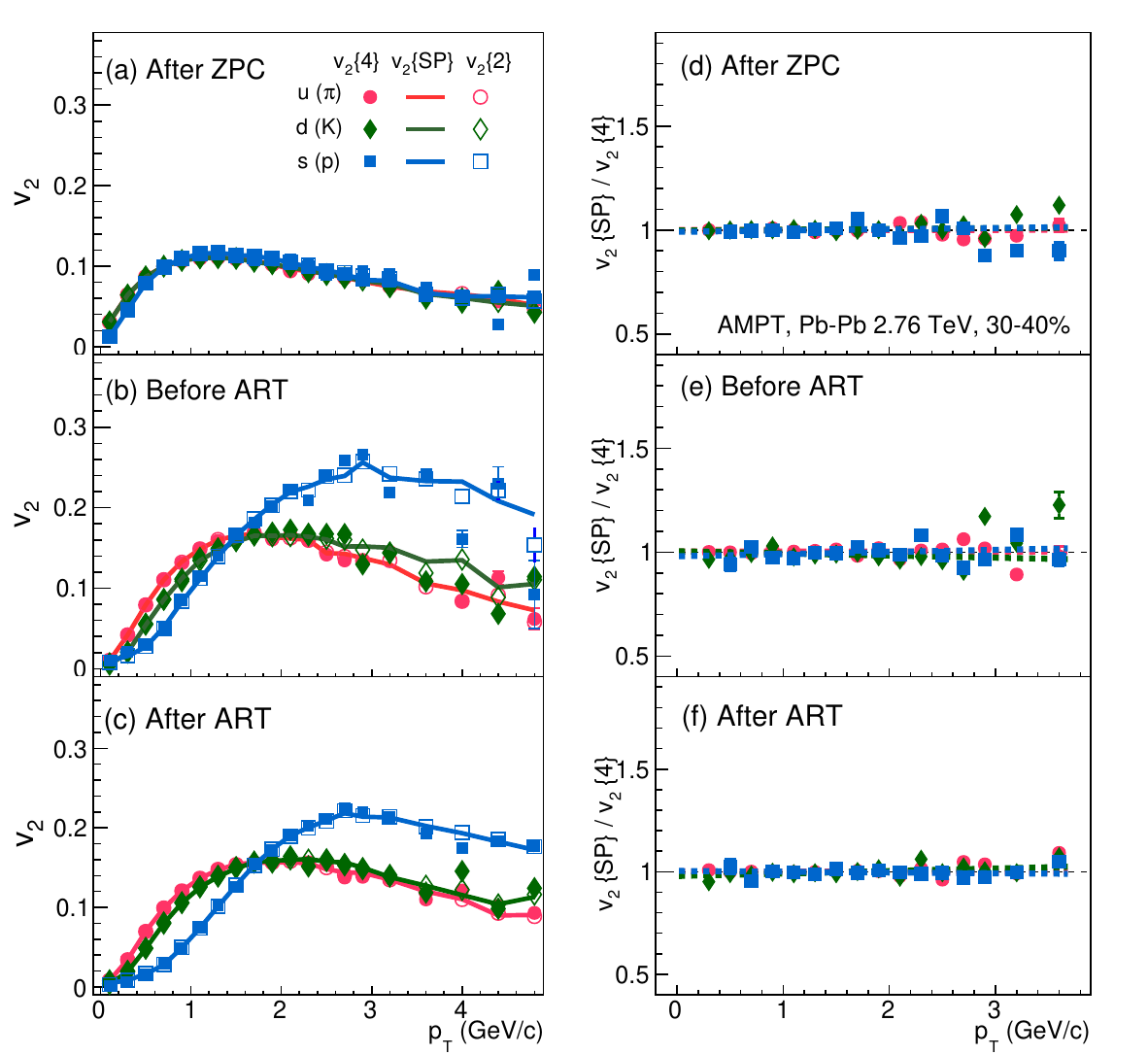}
\caption{(Color online) Same as Fig.~\ref{fig:fig_v2}, but with event mixing applied.}
\label{fig:fig_mix}
\end{figure*}

To investigate the complete evolution of flow fluctuations in AMPT, the analysis begins at the initial stage where partons are directly fragmented from HIJING before any interaction. Figure~\ref{fig:fig_zpct0} presents the \pt dependent $v_2$ for $u$, $d$, and $s$ quarks before ZPC in semi-central collisions. Significant deviations are observed between different flow measurement methods: \vtwosp remain consistent with zero at all \pt, while \vtwotwo exhibits positive values with a clear \pt dependence.
Although the initial-state partons exhibit positive spatial eccentricity, the absence of dynamical interactions at this stage prevents its conversion into momentum-space anisotropy. This physical picture is correctly captured by \vtwosp, which represents the genuine flow.
On the other hand, the observed positive \vtwotwo values do not represent genuine flow, but rather originate primarily from non-flow correlations dominated by mini-jet contributions inherited from the HIJING initial conditions, as evidenced by their characteristic \pt dependence - the magnitude increases with \pt, reflecting the growing mini-jet contribution at higher momenta.
We have also examined the \vtwofour values, but consistently observe opposite signs for the fourth-order cumulant ($c_4$) compared to expectations. Similar to \vtwotwo, the \vtwofour values at this initial stage contain no genuine flow information.

During partonic scattering, the initial geometric eccentricity is converted into momentum-space anisotropy through parton interactions, generating \vtwo. The underlying mechanism, governed by binary parton collisions with cross section $\sigma$, has been extensively studied in Refs.~\cite{ampt_evo1, ampt_evo2, ampt_evo3, ampt_evo4}.
Figure~\ref{fig:fig_v2} (a) displays the \pt-dependence of \vtwotwo, \vtwofour, and \vtwosp for $u$, $d$, and $s$ quarks after ZPC. All three observables exhibit positive values with reasonable \pt dependence. Notably, the expected hierarchy $\vtwotwo > \vtwosp > \vtwofour$ emerges clearly except at very low \pt, which is qualitatively consistent with the theoretical expectations from Eq.~(\ref{Eq:fluc}) as well as the ALICE measurement~\cite{alice_sp_flow, alice_sp_pidflow}.
The ratio \vtwosp/\vtwofour, shown in Fig.~\ref{fig:fig_v2} (d), rises monotonically from unity to $\approx 1.5$ with increasing \pt. This can be attributed to the intrinsic \pt-dependent flow fluctuations. In contrast, the ratio \vtwosp/\vtwotwo exhibits minimal \pt dependence, consistent with previous findings in Ref.~\cite{ampt_flow_fluc1}. This demonstrates that \vtwofour is more sensitive to \pt dependent flow fluctuations than \vtwotwo, as the latter could be masked by trivial correlations including non-flow effects.
It should be particularly noted that no quark flavor dependence is observed in any of three flow observables, and linear fits to the ratios yield comparable slopes and intercepts for all quark species. This confirms that $u$, $d$, and $s$ quarks undergo identical dynamical evolution during the flow-generating stage in ZPC.

\begin{figure*}
\centering
\includegraphics[width=.9\linewidth]{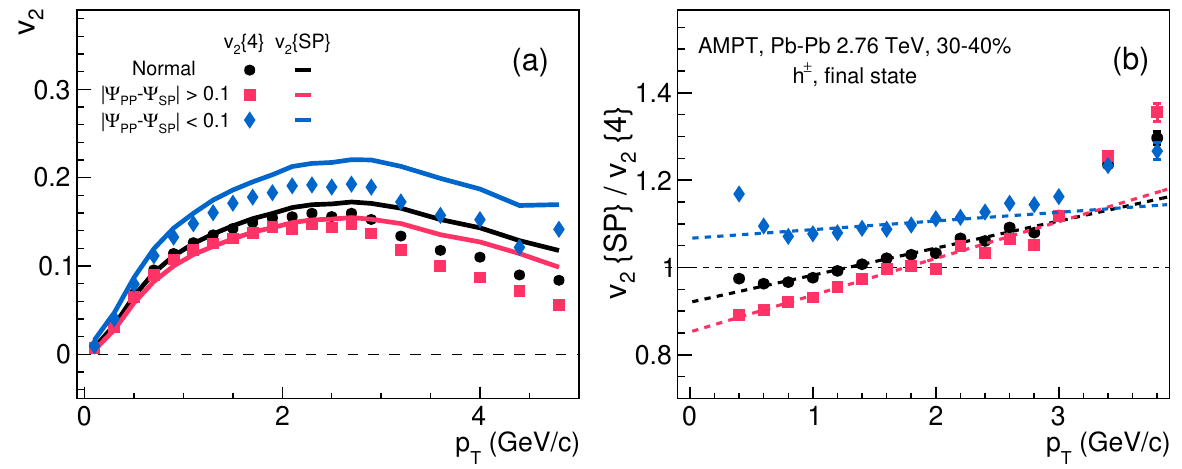}
\caption{(Color online) The \pt dependent $v_2 \{2\}$,  $v_2 \{4\}$ and $v_2 \{\rm SP\}$ and their ratios for inclusive charged hadrons in three $|\Psi_{\rm RP} - \Psi_{\rm PP}| $ conditions in 30-40\% AMPT Pb-Pb collisions.}
\label{fig:fig_psi}
\end{figure*}

Following the completion of partonic interactions, the coalescence mechanism combines nearby quarks into hadrons, further enhancing the momentum-space anisotropy. This process naturally gives rise to characteristic flow signatures, including mass ordering ($v_2^\pi > v_2^K > v_2^p$) below $\pt \approx 1.5$ GeV/$c$ and baryon-meson grouping ($v_2^p > v_2^\pi \approx v_2^K$) at higher \pt, as shown in Fig.~\ref{fig:fig_v2} (b). The hierarchy $\vtwotwo > \vtwosp > \vtwofour$ observed at the partonic level persists for all three hadron species. Figure~\ref{fig:fig_v2} (e) reveals that the ratio $\vtwosp/\vtwofour$ maintains its increasing trend with \pt, confirming the growing role of fluctuations at higher momenta. Notably, in contrast to the partonic stage, hadronic results show particle species dependent deviations in this ratio ($\pi \geq K > p$), matching ALICE measurements. By comparing panels (a) and (b), we conclude that the experimentally observed hierarchy emerges directly from coalescence dynamics.

The hadronic interaction stage in AMPT, which includes elastic scatterings and resonance decays, represents the final evolution step. These processes typically reduce the primordial momentum anisotropy. As shown in Fig.~\ref{fig:fig_v2} (c), the absolute $v_2$ values, particularly for protons, decrease slightly after ART, as expected, while the characteristic mass ordering ($v_2^\pi > v_2^K > v_2^p$ at low \pt) and baryon-meson splitting ($v_2^p > v_2^{\pi,K}$ at intermediate \pt) remain clearly visible.
Hadronic interactions also modify flow fluctuations (Fig.~\ref{fig:fig_v2} (f)): the ratio $\vtwosp/\vtwofour$ converges to nearly identical values for pions and kaons, while maintaining a systematic deviation for protons. This can be explained by the dominance of pion interactions and decays in ART. However, due to large experimental uncertainties in the corresponding measurements prevent a robust quantification of hadronic effects, highlighting the need for future precision studies.

\begin{figure*}
\centering
\includegraphics[width=.9\linewidth]{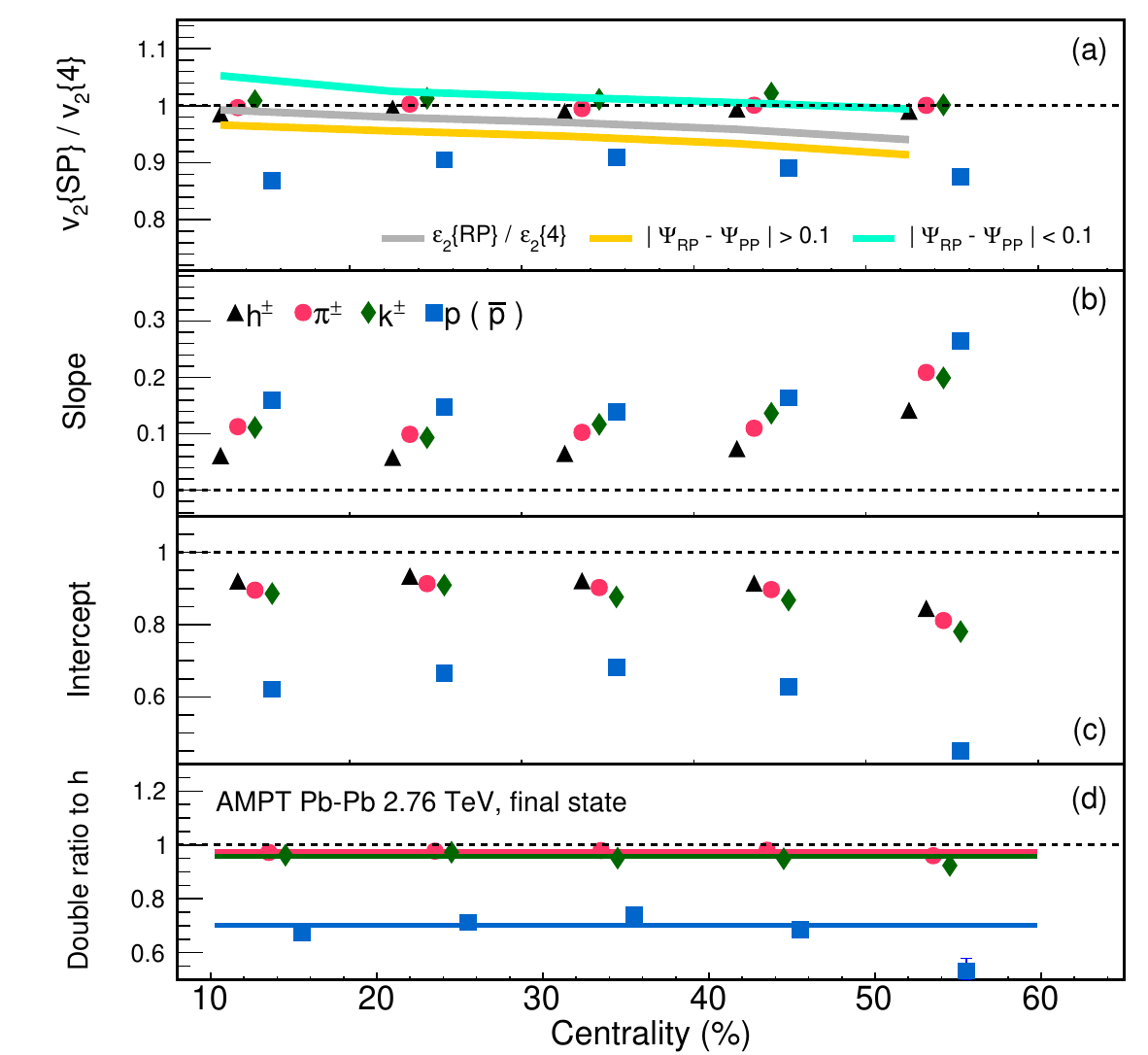}
\caption{(Color online) The centrality dependence of \vtwosp/\vtwofour, slope, intercept and double ratio to $h$ for $\pi, K, p$ in AMPT Pb-Pb collisions. The \vtwosp/\vtwofour is also compared with \etworp/\etwofour is panel (a).}
\label{fig:fig_cent}
\end{figure*}

\subsection{Crosscheck with event mixing and selected $\Psi_{\rm PP}$}

To further confirm the effect of flow fluctuations in single events, a mix-event method is applied to estimate the genuine flow. Mix-event method is widely used in various correlations studies~\cite{alice_corr1, alice_corr2}, which is able to eliminate all intrinsic intra-event correlations while preserving common dynamic features across an event ensemble. Usually, the mix-event method is inapplicable for flow analyses where reaction planes are totally randomized event-by-event. In this study, however, the initial reaction plane is always fixed along $x$-axis ($\Psi=0$) for all events, which enables the implementation of event mixing. Specifically, for $v_2 \{\rm SP\}$, since spectator planes in all events are required to be zero, the results simply remain unchanged; for $v_2 \{2\}$ and $\{4\}$, the particles of interest in the current event are paired with reference particles from other events within the same centrality interval.

Figure~\ref{fig:fig_mix} presents the results analogy to Fig.~\ref{fig:fig_v2} but with event mixing. Three vertical panels show results for stages after ZPC, before ART and after ART, respectively, while two horizontal panels display \pt dependent \vtwo and \vtwosp/\vtwofour ratio, respectively. Compared to Fig.~\ref{fig:fig_v2}, the \vtwosp remains identical across all evolution stages as expected. Notably, after event mixing, both \vtwotwo and \vtwofour converge to \vtwosp. This firmly demonstrates that the deviations among three flow observables observed in Fig.~\ref{fig:fig_v2} originate from single-event flow fluctuations.

An alternative approach to test flow fluctuations involves direct selection based on the participant plane $\Psi_{\rm PP}$. Initial fluctuations lead to event-by-event variations in parton distribution asymmetry within a given centrality class: events with more asymmetric distributions yield smaller \vtwo values, while those with more symmetric distributions produce larger \vtwo.
The results presented in previous sections are averaged measurements. To probe fluctuations, we categorize events by their participant plane orientation difference relative to the genuine plane $\Psi_{\rm RP}$, which follows a narrow normal distribution centered at zero. The event sample is divided into two subsets: $|\Psi_{\rm RP} - \Psi_{\rm PP}| < 0.1$ (more symmetric initial conditions and smaller fluctuations) and $|\Psi_{\rm RP} - \Psi_{\rm PP}| > 0.1$ (more asymmetric initial conditions and larger fluctuations).
Figure~\ref{fig:fig_psi} compares the \pt-dependent \vtwo between event selections based on participant plane alignment. As predicted, both $\vtwosp$ and $\vtwofour$ show systematically larger values for events with strong participant-spectator plane alignment ($|\Psi_{\rm RP} - \Psi_{\rm PP}| < 0.1$) compared to misaligned events ($|\Psi_{\rm RP} - \Psi_{\rm PP}| > 0.1$), with the unselected sample results lying between these two cases. The $\vtwosp/\vtwofour$ ratio exhibits corresponding variations: events with reduced fluctuations yield a flatter \pt dependence characterized by smaller slopes and larger intercepts, while enhanced fluctuations produce steeper trends and smaller intercepts. This test could be experimentally examined by correlating mid-rapidity TPC plane with forward ZDC plane, offering another probe of the relationship between flow fluctuations and the observed \pt-dependent $\vtwosp/\vtwofour$.

\subsection{Comparison with eccentricity and centrality dependence of slope, intercept and double ratio}

The elliptic flow originates from the initial spatial eccentricity of nucleons/partons, while flow fluctuation arise from intrinsic event-by-event asymmetry. Comparing the final-state flow ratio \vtwosp/\vtwofour with the initial-state eccentricity ratio \etworp/\etwofour helps to reveal the evolution of collective dynamics. 
Figure~\ref{fig:fig_cent} (a) shows the comparison of the \vtwo and \etwo ratios. The \vtwo ratios are extracted from Fig.~\ref{fig:fig_cent} over the full \pt range specified in Sec.~\ref{method}, while the \etwo ratios are calculated using Eq.~(\ref{Eq:e2}) with coordinates of all initial-state partons (prior to ZPC) without flavor distinction. For \vtwosp/\vtwofour, the particle species dependent hierarchy $\pi \simeq K > p$ can be seen across all centrality classes.
The calculated \etworp/\etwofour exhibits a consistent trend with quantitative values provided in Tab.~(\ref{tab:ecc}). When examining events selected by $\Psi_{\rm PP}$ alignment as described earlier, both ratios show correlated behavior: increasing for events with weaker fluctuations (smaller $\Psi_{\rm PP}$) and decreasing for those with stronger fluctuations (larger $\Psi_{\rm PP}$).
Notably, the AMPT calculated ratios, consistent with EPM and T$_{\rm R}$ENTo calculations, systematically underestimate the ALICE measurements~\cite{alice_sp_flow}. This discrepancy suggests that experimentally reconstructed spectator planes is less correlated with participant planes than theoretically modeled, thereby resulting in $\vtwosp/\vtwofour$ above unity in experimental data.
 
\begin{center}
\begin{table}[h]
\caption{The \etworp, \etwofour and their ratio in all centrality bins, calculated using Eq.~(\ref{Eq:e2}) with coordinates of all initial-state partons.}
\centering
\begin{tabular}{c | c | c | c}
Centrality (\%)& $\etworp$ & $\etwofour$ & $\etworp/\etwofour$ \\
\hline
10-20 & 0.1932 & 0.1946 & 0.9927 \\
\hline
20-30 & 0.2704 & 0.2760 & 0.9798 \\
\hline
30-40 & 0.3261 & 0.3359 & 0.9708 \\
\hline
40-50 & 0.3642 & 0.3799 & 0.9586 \\
\hline
50-60 & 0.3869 & 0.4113 & 0.9406 \\
\end{tabular}
\label{tab:ecc}
\end{table}
\end{center}

In the middle two panels (b) and (c) of Fig.~\ref{fig:fig_cent}, the centrality dependence of the slope and intercept parameters extracted from Fig.~\ref{fig:fig_v2} are presented, respectively.
The non-zero values of both parameters demonstrate significant non-linear coupling between the initial-state eccentricity and the developed elliptic flow. Both parameters also exhibit similar centrality dependence for pions, kaons, and protons with distinct particle-type splitting patterns. This reflects the combined effects of initial-state fluctuations and the hadronization of quark coalescence. Based on the conclusion derived from Fig.~\ref{fig:fig_psi}, protons is believed to experience stronger flow fluctuations than pions and kaons, as evidenced by their larger slopes and smaller intercepts.

To quantify the relative difference between $\pi$, $K$ and $p$, ALICE measured the intercepts ratios of identified hadrons to charged hadrons, which is supposed to effectively cancel the common linear hydrodynamic response contribution. 
As shown in panel (d), these $double~ratios$ exhibit a clear centrality-dependent hierarchy of $\pi \geq K > p$, with centrality-averaged values (uncertainties at the $\sim0.1\%$ level) of \\
\begin{equation} \nonumber
\pi:K:p = 0.98:0.96:0.70 .
\end{equation}
Such a hierarchy qualitatively agrees with ALICE result, resembling typical hydrodynamic ordering of identified particle flow. However, compared to the ALICE preliminary results (uncertainties at the $\sim1\%$ level) of 
\begin{equation} \nonumber
\pi:K:p = 1:0.98:0.96 ,
\end{equation}
the small difference between $\pi$ and $K$,  along with the relatively lower values for $p$, reflects limitations in AMPT's treatment of hadronic interactions, which should be further improved in future studies.

\section{Summary} \label{summary}

The ALICE experiment has recently measured the \pt dependent \vtwo fluctuations for charged pions, kaons, and (anti-)protons through the ratio \vtwosp/\vtwofour. Clear dependencies on both \pt and particle species have been observed, however, these features have yet to be systematically explored within hydrodynamic models, highlighting the need for further theoretical investigation. In this work, we conduct a phenomenological analysis using the AMPT model to trace the full evolution of particle species dependent \vtwo fluctuations.
We demonstrate that the flow fluctuations, as characterized by \vtwosp/\vtwofour, along with the development of \vtwo itself, originate during the partonic interaction stage. At this stage, \vtwosp/\vtwofour exhibits a clear linear dependence on \pt with a positive slope, showing no distinction among quark flavors. As quarks begin to coalesce into hadrons, this linear \pt dependence persists. Crucially, particle species dependence emerges, with the intercepts following the order $\pi > K > p$, consistent with ALICE observations. This strongly indicates that the observed particle species dependence stems directly from flow fluctuations intertwined with coalescence hadronization. Subsequent hadronic interactions in AMPT appear to reduce the difference between $\pi$ and $K$, while preserving the deviation for $p$.

We also show that in AMPT, the final-state \vtwosp/\vtwofour follows the initial-state \etworp/\etwofour in most centrality classes. The extracted slope and intercept from the linear \pt dependence of \vtwosp/\vtwofour exhibit a clear particle-species dependence. Furthermore, the double ratio of the intercepts for identified hadrons to those for inclusive hadrons reflects the same trend. These results qualitatively reproduce the ALICE measurements, and the reported values may serve as useful references for future experimental comparisons. 
We also point out that (1) \vtwosp/\vtwofour is more sensitive to both \pt and particle species than \vtwosp/\vtwotwo, making it a more effective observable for probing flow fluctuations; (2) by employing a novel event mixing technique, the difference among \vtwosp, \vtwotwo and \vtwofour vanish, and all converge to the genuine \vtwo, indicating that the observed flow fluctuations originate from the intra-event correlations; (3) selecting events based on the deviation of $\Psi_{\rm PP}$ from $\Psi_{\rm RP}$ allows for preferential sampling of events with stronger (weaker) flow fluctuations, which correspondingly leads to a steeper (shallower) slope and a smaller (larger) intercept in the \pt-\vtwosp/\vtwofour relation.

In the future, the contribution of hadronic effects should be further investigated to clarify the discrepancies with ALICE data. Given the small variation in \vtwo between 2.76 TeV and 5.02 -- 5.36 TeV~\cite{alice_flow_2tev, alice_flow_5tev}, our calculations and the underlying mechanism presented here are expected to remain valid across different LHC energies. With ALICE Run 3 currently collecting higher-statistics data, we anticipate future comparisons with more precise experimental results to provide deeper insights into the transport dynamics and hadronization processes of the QGP.

\section*{Acknowledgments}

We are grateful to Michael Rudolf Ciupek and Ilya Selyuzhenkov for their invaluable insights, which inspired this work, and for their enlightening discussions. We also thank Chen Zhong for his dedicated maintenance of the computing resources. This work is supported by the National Natural Science Foundation of China (Nos. 12322508, 12061141008, 12147101), the National Key Research and Development Program of China (No. 2024YFA1610802), the Science and Technology Commission of Shanghai Municipality (No. 23590780100) and the 111 Project (MOEC, MSTC).



\begin{thebibliography} {99}
\bibitem{alice_qgp} S. Acharya et al., The ALICE experiment: a journey through QCD, Eur. Phys. J. C 84, 813 (2024).
\bibitem{nst_alice} Q.-Y. Shou et al., Properties of QCD matter: a review of selected results from ALICE experiment, NUCL SCI TECH 35, 219 (2024).
\bibitem{nst_star} J.-H. Chen et al., Properties of the QCD matter: review of selected results from the relativistic heavy ion collider beam energy scan (RHIC BES) program, NUCL SCI TECH 35, 214 (2024).
\bibitem{flow_review1} U. Heinz and R. Snellings, Collective Flow and Viscosity in Relativistic Heavy-Ion Collisions, Annu. Rev. Nucl. Part. Sci. 63, 123 (2013).
\bibitem{ecc1} R. S. Bhalerao and J.-Y. Ollitrault, Eccentricity fluctuations and elliptic flow at RHIC, Physics Letters B 641, 260 (2006).
\bibitem{ecc2} J.-Y. Ollitrault, A. M. Poskanzer, and S. A. Voloshin, Effect of flow fluctuations and nonflow on elliptic flow methods, Phys. Rev. C 80, 014904 (2009).
\bibitem{ecc3} B. Alver et al., Importance of correlations and fluctuations on the initial source eccentricity in high-energy nucleus-nucleus collisions, Phys. Rev. C 77, 014906 (2008).
\bibitem{flow_ep} A. M. Poskanzer and S. A. Voloshin, Methods for analyzing anisotropic flow in relativistic nuclear collisions, Phys. Rev. C 58, 1671 (1998).
\bibitem{flow_cumu1} A. Bilandzic, R. Snellings, and S. Voloshin, Flow analysis with cumulants: Direct calculations, Phys. Rev. C 83, 044913 (2011).
\bibitem{flow_cumu2} A. Bilandzic, C. H. Christensen, K. Gulbrandsen, A. Hansen, and Y. Zhou, Generic framework for anisotropic flow analyses with multiparticle azimuthal correlations, Phys. Rev. C 89, 064904 (2014).
\bibitem{fluc1} H. Holopainen, Event-by-event hydrodynamics and elliptic flow from fluctuating initial states, Phys. Rev. C 83, (2011).
\bibitem{star_flow_fluc1}] G. Agakishiev et al., Energy and system-size dependence of two- and four-particle v 2 measurements in heavy-ion collisions at $\sqrt{{s}_{\rm NN}}$ = 62.4 and 200 GeV and their implications on flow fluctuations and nonflow, Phys. Rev. C 86, 014904 (2012).
\bibitem{alice_sp_flow} S. Acharya et al., Elliptic flow of charged particles at midrapidity relative to the spectator plane in Pb–Pb and Xe–Xe collisions, Physics Letters B 846, 137453 (2023).
\bibitem{alice_xe_flow} S. Acharya et al., Anisotropic flow in Xe–Xe collisions at $\sqrt{{s}_{\rm NN}}=5.44$ TeV, Physics Letters B 784, 82 (2018).
\bibitem{trento} Reduced Thickness Event-by-event Nuclear Topology ($\rm T_{R}ENTo$), \url{http://qcd.phy.duke.edu/trento}.
\bibitem{epm} L. Yan, J.-Y. Ollitrault, and A. M. Poskanzer, Eccentricity distributions in nucleus-nucleus collisions, Phys. Rev. C 90, 024903 (2014).
\bibitem{alice_sp_pidflow} M. R. Ciupek (for the ALICE Collaboration), talk given at Initial Stages 2023, \url{https://indico.cern.ch/event/1043736/contributions/5363772}.
\bibitem{ampt_flow_fluc1} N. Magdy, X. Sun, Z. Ye, O. Evdokimov, and R. Lacey, Investigation of the Elliptic Flow Fluctuations of the Identified Particles Using the a Multi-Phase Transport Model, Universe 6, 9 (2020).
\bibitem{ampt1} Z.-W. Lin, C. M. Ko, B.-A. Li, B. Zhang, and S. Pal, Multiphase transport model for relativistic heavy ion collisions, Phys. Rev. C 72, 064901 (2005).
\bibitem{ampt2} Z.-W. Lin and L. Zheng, Further developments of a multi-phase transport model for relativistic nuclear collisions, NUCL SCI TECH 32, 113 (2021).
\bibitem{ampt_flow1} A. Bzdak and G.-L. Ma, Elliptic and Triangular Flow in p-Pb and Peripheral Pb-Pb Collisions from Parton Scatterings, Phys. Rev. Lett. 113, 252301 (2014).
\bibitem{ampt_flow2} L. Ma, G. L. Ma, and Y. G. Ma, Initial partonic eccentricity fluctuations in a multiphase transport model, Phys. Rev. C 94, 044915 (2016).
\bibitem{ampt_flow3} L. Zheng, H. Li, H. Qin, Q.-Y. Shou, and Z.-B. Yin, Investigating the NCQ scaling of elliptic flow at LHC with a multiphase transport model, Eur. Phys. J. A 53, 124 (2017).
\bibitem{ampt_flow4} S.-Y. Tang, L. Zheng, X.-M. Zhang, and R.-Z. Wan, Investigating the elliptic anisotropy of identified particles in p–Pb collisions with a multi-phase transport model, NUCL SCI TECH 35, 32 (2024).
\bibitem{hijing} M. Gyulassy and X.-N. Wang, HIJING 1.0: A Monte Carlo program for parton and particle production in high energy hadronic and nuclear collisions, Computer Physics Communications 83, 307 (1994). \newline
X.-N. Wang and M. Gyulassy, hijing: A Monte Carlo model for multiple jet production in pp, pA, and AA collisions, Phys. Rev. D 44, 3501 (1991).
\bibitem{zpc} B. Zhang, ZPC 1.0.1: a parton cascade for ultrarelativistic heavy ion collisions, Computer Physics Communications 109, 193 (1998).
\bibitem{coal} R. Fries, V. Greco, and P. Sorensen, Coalescence Models for Hadron Formation from Quark-Gluon Plasma, Annu. Rev. Nucl. Part. Sci. 58, 177 (2008).
\bibitem{art} B.-A. Li, Formation of superdense hadronic matter in high energy heavy-ion collisions, Phys. Rev. C 52, 2037 (1995). \newline
B.-A. Li, A. T. Sustich, B. Zhang, and C. M. Ko, Studies of superdense hadronic matter in a relativistic transport model, Int. J. Mod. Phys. E 10, 267 (2001).
\bibitem{ampt_par1} J. Xu and C. M. Ko, Pb-Pb collisions at $\sqrt{{s}_{\rm NN}}=2.76$ TeV in a multiphase transport model, Phys. Rev. C 83, 034904 (2011).
\bibitem{ampt_par2} J. Xu and C. M. Ko, Triangular flow in heavy ion collisions in a multiphase transport model, Phys. Rev. C 84, 014903 (2011).
\bibitem{ampt_cme_sp} B.-X. Chen, X.-L. Zhao, and G.-L. Ma, Difference between signal and background of the chiral magnetic effect relative to spectator and participant planes in isobar collisions at $\sqrt{{s}_{\rm NN}}=200$ GeV, Phys. Rev. C 109, 024909 (2024).
\bibitem{ampt_evo1} L. He, T. Edmonds, Z.-W. Lin, F. Liu, D. Molnar, and F. Wang, Anisotropic parton escape is the dominant source of azimuthal anisotropy in transport models, Physics Letters B 753, 506 (2016).
\bibitem{ampt_evo2} H. Li, L. He, Z.-W. Lin, D. Molnar, F. Wang, and W. Xie, Origin of the mass splitting of elliptic anisotropy in a multiphase transport model, Phys. Rev. C 93, 051901 (2016).
\bibitem{ampt_evo3} G.-L. Ma and A. Bzdak, Flow in small systems from parton scatterings, Nuclear Physics A 956, 745 (2016).
\bibitem{ampt_evo4} Z.-W. Lin, L. He, T. Edmonds, F. Liu, D. Molnar, and F. Wang, Elliptic Anisotropy v2 May Be Dominated by Particle Escape instead of Hydrodynamic Flow, Nuclear Physics A 956, 316 (2016).
\bibitem{alice_corr1} B. Abelev et al., Long-range angular correlations of $\pi$, K and p in p–Pb collisions at $\sqrt{{s}_{\rm NN}}=5.02$ TeV, Physics Letters B 726, 164 (2013).
\bibitem{alice_corr2} J. Adam et al., Insight into particle production mechanisms via angular correlations of identified particles in pp collisions at $\sqrt{{s}}=7$ TeV, Eur. Phys. J. C 77, 569 (2017).
\bibitem{alice_flow_2tev} The ALICE collaboration et al., Elliptic flow of identified hadrons in Pb-Pb collisions at $\sqrt{{s}_{\rm NN}}=2.76$ TeV, J. High Energ. Phys. 2015, 190 (2015).
\bibitem{alice_flow_5tev} The ALICE collaboration et al., Anisotropic flow and flow fluctuations of identified hadrons in Pb–Pb collisions at $\sqrt{{s}_{\rm NN}}=5.02$ TeV, J. High Energ. Phys. 2023, 243 (2023).
\end{thebibliography}
\end{document}